\documentstyle[aas2pp4]{article}
\begin{document}

\title{A New Method For Galaxy Cluster Detection I: The Algorithm}
\author{Michael D. Gladders}
\author{{\it and}}
\author{H.K.C. Yee}

\vspace{2.0cm}
\affil{Department of Astronomy, University of Toronto, 60 St. George
  St., Toronto, ON, M5S 3H8, Canada}
\authoremail{gladders,hyee@astro.utoronto.ca}

\begin{abstract}
  Numerous methods for finding clusters  at moderate to high redshifts
  have   been proposed in recent   years,  at wavelengths ranging from
  radio  to X-rays.   In  this  paper we  describe  a new  method  for
  detecting clusters in two-band  optical/near-IR  imaging data.   The
  method relies upon the observation  that all  rich clusters, at  all
  redshifts observed     so far, appear   to have   a red  sequence of
  early-type galaxies.  The emerging picture is that all rich clusters
  contain a core population of  passively evolving elliptical galaxies
  which  are coeval and formed  at high redshifts. The proposed search
  method exploits this strong empirical fact by using the red sequence
  as a direct indicator of  overdensity.  The fundamental advantage of
  this approach is that  with appropriate filters, cluster  elliptical
  galaxies at a given redshift are  redder than all normal galaxies at
  lower  redshifts.  A simple color cut  thus virtually eliminates all
  foreground contamination,  even  at significant redshifts.   In this
  paper,   one of   a  series of    two, we   describe the  underlying
  assumptions and   basic  techniques of  the method   in  detail, and
  contrast the method with those used by other  authors.  We provide a
  brief demonstration of  the effectiveness of  the  technique using a
  real photometric sample with redshift  data, and from this  conclude
  that the method offers a powerful yet simple  way of identify galaxy
  clusters. We find that the method  can reliably detect structures to
  masses as   small as   groups  with velocity   dispersions  of  only
  $\sim300{\rm\thinspace    km}{\rm\thinspace   s}^{-1}$, with
  redshifts  for all detected  structures estimated  to an accuray  of
  $\sim$10\%.

\end{abstract}


\section{Introduction}
The detection  and characterization  of rich  clusters  of galaxies from
low ($z\sim0$)  to  high ($z\gtrsim1.0$)  redshifts provides  a
crucial     test     of     both    cosmological    models      
(e.g., 
\markcite{carlberg1}Carlberg         et       al.                1996;
\markcite{fan}Fan, Bahcall, \& Cen 1997;
\markcite{oukbir}Oukbir, Bartlett \&  Blanchard  1997;
\markcite{eke}Eke et al. 1998; 
\markcite{borgani}Borgani et al. 1999;
\markcite{ledlow}Ledlow et al.   1999;
\markcite{reichart}Reichart et al. 1999;
\markcite{viana}Viana \& Liddle 1999;
\markcite{henry4}Henry 2000, 
to list a recent subset of an extensive lterature)
and galaxy
evolution  
(see \S3 for an extensive discussion).
Clusters
trace  structure   in    the   Universe   to  large    scales   
(e.g., 
\markcite{bahcall2}Bahcall \& Soneira 1983;
\markcite{gramann}Gramann  et  al.      1995;
\markcite{tadros}Tadros, Efstathiou, \&  Dalton 1998;
\markcite{borgani2}Borgani, Plionis \& Kolokotronis 1999
), 
and provide many  examples of both strong
and  weak   lensing (see \markcite{mellier}Mellier 1999 for a recent review).
The obvious importance of
galaxy clusters  has led to numerous galaxy  cluster surveys in recent
years, both optically 
(e.g.,
\markcite{gho}Gunn, Hoessel \& Oke 1986;
\markcite{abell89}Abell, Corwin \& Olowin 1989;
\markcite{couch}Couch et al. 1991
\markcite{lumsden}Lumsden et al. 1992;
\markcite{lidman}Lidman \& Peterson 1996;
\markcite{post}Postman et al. 1996;
\markcite{dalton}Dalton et al. 1997;
\markcite{zaritsky}Zaritsky et al. 1997;
\markcite{ostrander}Ostrander et al. 1998;
\markcite{scodeggio}Scodeggio et al. 1999;
\markcite{gal}Gal et al. 2000
) 
and with X-rays (\markcite{gioia}Gioia \& Luppino 1994;
\markcite{scharf1}Scharf et al. 1997;
\markcite{rosati}Rosati  et al. 1998;
\markcite{vikh}Vikhlinin et al. 1998;
\markcite{degrandi}de Grandi et al. 1999;
\markcite{romer}Romer et al. 2000).  As pointed out by
\markcite{kepner}  Kepner  et al.  (1999),  each  of these surveys has
defined its  own  detection algorithm,  considered  by  the individual
authors to  be appropriate to  their particular data. In essence, each
of these detection algorithms makes assumptions about ``what a cluster
looks like''  and then searches the  relevant data  to identify likely
cluster locations.

\paragraph{}
All cluster finding methods suffer from selection effects. At minimum,
each will be biased against any clusters which do not fit the
particular cluster definition used. Even if cluster samples appear
essentially congruent at different wavelengths, scatter in the
individual properties of individual clusters will still engender
somewhat different samples if surveys are done at different
wavelengths. For example, results from the CNOC1 survey show that the
optical richness (important for optically selected catalogs) and X-ray
luminosity (important for X-ray selected catalogs) of a sample of 16
X-ray selected clusters at $0.1<z<0.55$ are only moderately correlated
(\markcite{cnoc1global}Yee et al. 2000a).  X-ray surveys, which
ultimately define clusters as diffuse X-ray sources, are potentially
biased against gas-poor clusters, and clusters in which the gas
distribution is compact and hence unresolved.  There is some limited
evidence for optically rich and massive, but X-ray weak clusters.  For
example, \markcite{bower1}Bower et al. (1994) found that a sample of
14 optically selected rich clusters at $z\sim0.4$ showed relatively
weak X-ray emission, though the significance of this result is
dependent on the interpretation of masses from minimal spectroscopic
data. \markcite{castander}Castander et al.  (1994) found similar results for
a sample of optically selected clusters at $z\sim0.8$. Optical surveys are
potentially biased against optically dark clusters -- deep
cluster-sized potential wells which, for whatever reason, have failed
to form a significant galaxy population.  Notably, there is little
evidence for many such clusters.  The X-ray selected sample of
\markcite{vikh} Vikhlinin et al. (1998) showed no evidence for any
optically dark clusters, with a sample of $\sim$ 200 X-ray selected
clusters from $z=0.015$ to $z>0.5$. A similar result is found in the
Bright SHARC sample (\markcite{romer}Romer et al. 2000). At higher
redshift, the cluster AXJ2019+112 at $z=1.01$, claimed by
\markcite{hattori}Hattori et al.  (1997) to be a dark cluster, has
been shown by \markcite{benitez} Ben\'{\i}tez et al.  (1999) to not be
dark at all. Clusters which are under-luminous in galaxian light may
exist, but appear to be the exception rather than the norm. Another
low-mass cluster type likely missed by optical techniques are the
so-called fossil groups (\markcite{ponman2}Ponman et al. 1994;
\markcite{vihk2}Vikhlinin et al. 1999; \markcite{romer}Romer et
al. 2000) which appear to be relatively common though not massive
(\markcite{vihk2}Vikhlinin et al. 1999). 

\paragraph{}
More complicated redshift dependent biases and uncertainties also
exist in most cluster catalogs, quite apart from the dominant
cosmological effects of redshift.  For example, optical-IR selected
samples are subject to the effects of cluster and field galaxy
evolution, which are, in general, not well determined at $z>1$.
Note, however, that the evolution of the {\it elliptical} galaxy
population in clusters has been shown by numerous authors (e.g.,
\markcite{stanford98} Stanford, Eisenhardt, \& Dickinson 1998 and 
references therein, hereafter SED98) to be remarkably simple and
homogeneous.  The apparent stability and smooth evolution of this
dominant cluster population implies the selection functions for
clusters defined by these galaxies can be computed with some
confidence, even at $z>1$.  For a given cluster finding technique to
be most useful, the reliable computation of selection functions is as
critical as the sensitivity of the algorithm (\markcite{post}Postman
et al. 1996).

\paragraph{}
The general selection functions for clusters in most surveys are
typically quite complicated functions of numerous parameters,
including, but not limited to, the cluster redshift and the survey
flux limits.  However, in most cases, the detection rate for clusters
of a given parameter set can generally be computed directly from the
survey data, by inserting and attempting to recover fiducial test
clusters. So long as the test clusters are a realistic representation
of the cluster populations, this procedure will produce reliable
results, and such techniques have been used successfully in several
cluster surveys (e.g.,
\markcite{post}Postman et al. 1996;
\markcite{ostrander}Ostrander et al. 1998;
\markcite{bramel}Bramel, Nichol \& Pope 2000). 
False positive detections are however more difficult to quantify,
as a thorough understanding of such false cluster detections requires
some $a$ $priori$ knowledge of the contaminants -- such as projection
in optical catalogs (e.g., \markcite{post}Postman et al. 1996) or
contamination by active galactic nuclei in X-ray selected catalogs
(e.g., \markcite{vikh} Vikhlinin et al. 1998).  Current attempts to
quantify the false positive contaminations rates in samples in the
absence of detailed follow-up data are typically based on some form of
resampling strategy (e.g., \markcite{bramel}Bramel, Nichol \& Pope
2000).

\paragraph{}
Indeed, the dominant cause of false-positive cluster candidates in
optical cluster catalogs is projection.  It has long been suggested
that there are significant projection effects in the Abell catalog
(e.g., \markcite{lucey}Lucey 1983), though the exact degree of
contamination is a matter of some debate (Collins et al. 1995). Even
more modern cluster catalogs suffer significant false cluster
detections due to random projections (\markcite{post}Postman et
al. 1996; \markcite{oke1}Oke, Postman, \& Lubin 1998;
\markcite{holden}Holden et al.  1999).  The projection-induced
contamination in the Abell catalog has been studied with exhaustive
n-body simulations, which show this to be primarily an artifact of the
large aperture used to initially define the Abell catalog
(\markcite{vanhaarlem}van Haarlem, Frenk, \& White 1997). These n-body
results demonstrate that the false-positive rates due to projection
for simulated cluster catalogs detected by either X-rays or 2-D galaxy
overdensities is in fact quite similar, provided a similar detection
aperture is used (\markcite{vanhaarlem}van Haarlem, Frenk, \& White
1997; also see \markcite{omaryee}L\'{o}pez-Cruz \& Yee 2000a,
hereafter LCY00).

\paragraph{}
Given the extreme importance of galaxy clusters for a variety of
studies, it is clear that large, well-understood samples of clusters
are needed. In such a setting, it desirable to have samples which are
selected in as many ways as possible, at as many wavelengths as
possible, as each sample will probe a somewhat different population
with somewhat different selection effects.  Furthermore, the cluster
finding algorithms used should be as efficient as possible in terms of
the required data, since clusters are rare objects and thus only found
in large numbers in observationally expensive wide-field surveys. The
method proposed below addresses these requirements. The method, which
we term the Cluster-Red-Sequence (CRS) method, exploits the
observational fact that the bulk of the early-type galaxies in all
rich clusters lie along a linear color-magnitude relation. This
relation, referred to hereafter as the red sequence, has been shown to
have an extremely small scatter (e.g., \markcite{bowera}
\markcite{bowerb} Bower, Lucey,
\& Ellis 1992b) and appears to be extremely homogeneous from cluster
to cluster  (LCY00).  In the  CRS method  clusters  are  detected  as
over-densities  in  projected  angular position,  color  and magnitude
simultaneously. The  color  of  the  red sequence provides   a precise
redshift estimator for  the detected  clusters.  The color  constraint
makes the CRS method  extremely insensitive to projection effects,  as
random projections do not exhibit the necessary red sequence signature
in the color-magnitude plane.  This makes the CRS method fundamentally
different from previous optical cluster finding algorithms.
The concept of using the red sequence as a cluster marker, or to
delineate cluster superstructures, has some history in the literature
(e.g., 
\markcite{omarthesis}Lopez-Cruz 1997;
\markcite{cfht1}Gladders \& Yee 1998;
\markcite{nick3}Kaiser et al. 1998;
\markcite{pasedena}Yee, Gladders, \& L\'{o}pez-Cruz 1999;
\markcite{lubin3}Lubin et al. 2000;
\markcite{omaryee}LCY00), 
and other cluster-finding techniques have been suggested which exploit
the dominance of early-type galaxies in clusters in some way
(\markcite{ostrander}Ostrander et al. 1998). However, the critical
point of the technique suggested here is the recognition that the
cluster red sequence can be effectively isolated from survey data with
only two filters, so long as the filter pair samples the
4000\AA\thinspace break. This technique precludes the need for
many-band imaging, such as is required for full photometric redshift
analyses (\markcite{kodama2}Kodama, Bell \& Bower 1999) and so is
efficient while still exploiting the available color information.
This paper represents the first effort to cast the CRS method as a
well-defined, well-motivated algorithm. The purpose is to describe an
algorithm which is amenable to stringent testing, and useful for
finding clusters in a wide range of optical-IR imaging data.

\paragraph{}
We have begun a large observational program, the Red-Sequence Cluster
Survey (RCS), to identify and characterize clusters using the CRS
method.  This paper, the first of a series of two, describes the basic
observational motivation for the CRS method, and provides a detailed
algorithm and a test of that algorithm using existing data with
spectroscopic redshifts.  The second paper in the series discusses the
CRS method in the specific context of the RCS dataset, focussing on a
detailed derivation of the RCS selection functions for clusters using
extensive simulations.

\paragraph{}
This paper is arranged as follows.  We describe and motivate the CRS
method in the context of previous work and current observational
constraints in \S2 and \S3.  A detailed implementation of the method is given
in \S4.  In \S5 we provide a simple but powerful test of the method,
using real redshift data from the CNOC2 Redshift Survey
(\markcite{yeecnoc2}Yee et al.  2000b).  Section 6 provides a brief
discussion of possible refinements to the basic method outlined
here. We summarize our findings in  \S7.  A cosmology of $H_0$=70
\hbox{${\rm\thinspace km}{\rm\thinspace s}^{-1}{\rm\thinspace Mpc
}^{-1}$}, $\Omega_M$=0.2 and $\Omega_\lambda$=0.0 is used throughout.

\section{The CRS Method}
By definition,  optical/IR  cluster-finding techniques  all rely  upon
overdensities in  galaxy distributions   as  the signature   of   mass
overdensities.  The basic   premise shared by  all  such techniques is
that galaxies are a reliable (though not necessarily unbiased, \markcite{nick2}Kaiser
1984) tracer    of  the underlying mass    distribution,  which is the
cosmologically interesting distribution.  So long as the mass-to-light
ratio    of  galaxy  clusters is  roughly    constant, or  evolving in
well-understood ways, this basic  premise seems sound. The most likely
complication is the possible   presence of dark clusters, though   the
current evidence indicates that such objects are rare at best (see \S1
above). Given   this  basic premise,  the  differences between cluster
finding-techniques  are based in the manner  in which each attempts to
identify the real  three-dimensional galaxy overdensities in typically
two-dimensional imaging data.  In  general, the more modern techniques
(the   matched-filter  technique,  Postman  et  al.   1996;  Kepner et
al. 1999) are  designed to be more sensitive  to true clusters  (while
being less sensitive to random  projections) at the expense of greater
susceptability to modeling  uncertainties.  The method we  propose and
test here makes  the  basic  assumption that   the presence of  a  red
sequence of  early-type  galaxies is  a  ubiquitous and near-universal
signature  of clusters.  This  relatively   strong assumption is  well
supported  by the current observational evidence,  and provides a very
efficient and powerful method for detecting clusters. In the following
we  describe the basic advantages of  this  method and discuss some of
the possible uncertainties, deferring   a detailed description of  the
algorithm to \S4.

\paragraph{}
The CRS method is motivated by the observation that all rich clusters
have a population of early type galaxies which follow a strict
color-magnitude relation.  An example of this is provided in Figure 1,
which shows the measured color-magnitude diagram for the $z=0.231$
cluster Abell 2390, from the HST data in Gladders et al. (1998). Even
in this field-contaminated data, the cluster red sequence is easily
recognized. Current evidence (discussed in detail in \S3) indicates
that this galaxy population is remarkably homogeneous, both within
individual clusters and between clusters. Moreover, it appears that
the stellar population which makes up the red sequence is formed at
high redshifts ($z_f > 2$).  In cold-dark-matter dominated scenarios
of hierarchical structure formation this is the expected result, as
present-day clusters are correlated with the most extreme initial
over-densities, which are the first to collapse. Overall, cluster
elliptical galaxies appear to have the oldest and most homogeneous
stellar populations, and so can be expected to be a relatively stable
marker of clustering.

\begin{figure}[h]
\figurenum{1}
\plotfiddle{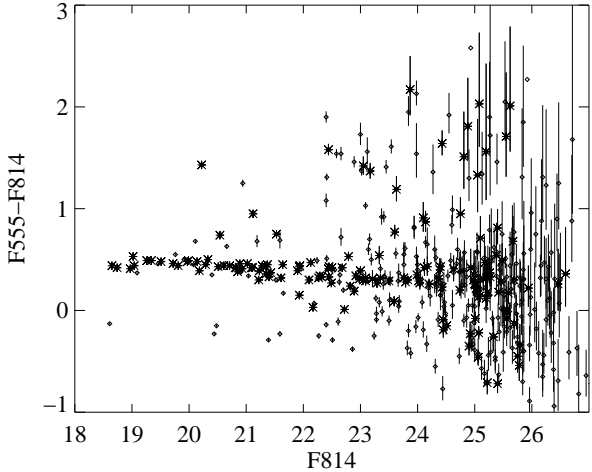}{6.5cm}{0.0}{50}{50}{-130}{0}
\caption{
The observed color-magnitude diagram for Abell 2390,
based on two-filter HST imaging of the cluster core. The data are from
Gladders et al. (1998). The asterisks indicate galaxies morphologically
selected as early-types, and diamonds indicate other galaxies in the
image. Error bars are 1-sigma.}
\end{figure}

\paragraph{} 
There are numerous observational reasons which make the red sequence
an attractive target for cluster finding, apart from its apparent
homogeneity. First, elliptical galaxies generally dominate the bright
end of the cluster luminosity function (e.g.,
\markcite{sandage}Sandage, Bingelli, \& Tamman 1985;
\markcite{thompson}Thompson 1986; \markcite{barger}Barger et al. 1998
-- though see
\markcite{rakos1}Rakos, Odell, \& Schombert [1997] for an exception to
this), and so are the most readily seen galaxies in a flux limited
survey.  Cluster elliptical galaxies are more luminous at higher
redshifts (e.g., \markcite{schade}Schade, Barrientos, \& Lopez-Cruz
1997; \markcite{vandokkum1}van Dokkum et al.  1998), consistent with
the age-induced fading expected in a passively evolving stellar
population.  In addition, the radial distribution of elliptical
galaxies in regular, centrally concentrated clusters is more compact
than that of other morphological types due to the morphology-density
relation (\markcite{dressler}Dressler et al.  1997 and references
therein), and so presents a higher contrast against the background.
Even in irregular clusters, with no well defined center, the
morphology density relation holds, though at lesser significance
(\markcite{dressler}Dressler et al.  1997), and elliptical galaxies
still trace the densest cluster regions.  Additionally, elliptical
galaxies have core-dominated compact brightness profiles, and so can
be morphologically selected with a high level of confidence
(\markcite{bobmorph}Abraham et al.  1994;
\markcite{gladders1}Gladders et al. 1998)

\begin{figure}[h]
\figurenum{2}
\plotfiddle{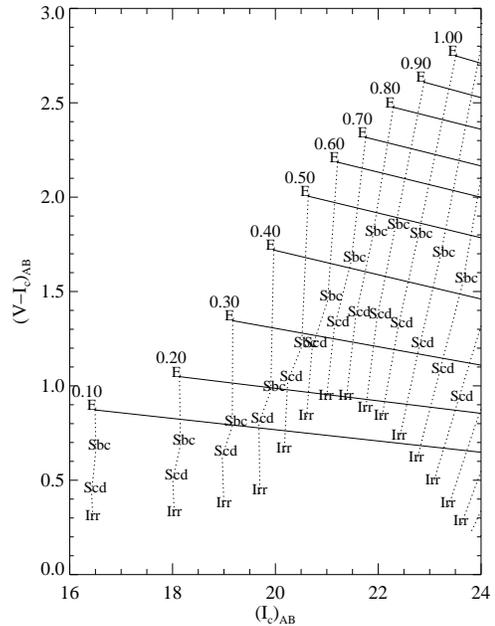}{8.5cm}{0.0}{40}{40}{-110}{0}
\caption{
A simulated (V-I$_c$)$_{AB}$ vs.  (I$_c$)$_{AB}$
color-magnitude diagram. Model apparent magnitudes and colors at
various redshifts for several types of galaxies at a fixed M$_I$ of
-22. The dotted lines connect galaxies at the same redshift.  Solid
near-horizontal lines show the expected slope of the red sequence at
each redshift. }
\end{figure}
 
\paragraph{}
An examination of Figure 2 reveals the final, critical advantage of
using elliptical galaxies to find clusters. Figure 2 shows model
color-magnitude tracks, over a range of redshift, for galaxies of
various spectral types (from \markcite{cww}Coleman, Wu, \& Weedman
1980), as well as the expected location of the cluster red sequence,
at similar $z$, using red sequence slopes from Kodama (1997).
Since cluster elliptical galaxies likely represent the oldest stellar
populations in the Universe, they are as red or redder than any other
galaxies at a given redshift.  Moreover, with properly chosen filters
straddling the 4000\AA\thinspace break, the cluster red sequence is as
red or redder than other galaxies at a given redshift {\it and all
lower redshifts}. As can be seen in Figure 2, this means that the bulk
of the contaminant galaxies within a color slice corresponding to a red
sequence at a given redshift are intrinsically bluer galaxies at yet
higher redshifts. {\it In other words, the CRS method does not
accumulate an overburden of foreground contamination when extended to
higher redshifts, effectively solving the critical foreground
projection problem with optical/IR cluster searches at such
redshifts.} Moreover, projection effects due to background structures
at higher redshift are minimized, because such structures will
generally be most significant at yet redder colors.

\paragraph{}
A secondary advantage of using the red sequence as a cluster tracer is
that the red sequence color provides an extremely precise redshift
indicator.  LCY00 noted that the red sequence $B-R$ (at a fixed
apparent $R$) versus  redshift diagram for low $z$ clusters shows a
remarkably small scatter of $\Delta z=0.008$ about the mean relation.
This had initially led
\markcite{omarthesis}L\'{o}pez-Cruz  (1997)   to   suggest that  the   red
sequence color would be a valuable redshift indicator.
\markcite{smail}Smail et al.
(1998) came to similar conclusions, though with a much larger
uncertainty, for their cluster sample at $z\sim0.25$.
\markcite{stanford98}SED98 found the red sequence color to be well
correlated to redshift to $z\sim0.9$.  Notably, the redshift scatter
implied by these simple photometric methods is comparable to or better
than that seen in four (or more) band photometric redshift studies
with calibration redshifts (\markcite{brunner}Brunner et al.  1997).
This precision is a result of the significant reduction in photometric
errors due to using many galaxies for a single color estimate, and the
intrinsic similarity of red sequence galaxies. Notably, this redshift
accuracy is significantly better than that achieved by the
matched-filter method (\markcite{holden1}Holden et al. 2000), because
the color-based redshift relies upon measuring the centroid of a peak
in the colour distribution whereas the matched-filter redshift
requires a measure of the cluster M$^*$. The former can be measured
with much greater precision than the latter.

\paragraph{}
In  brief, the CRS method  defines a cluster  as an overdensity on the
sky  which also demonstrates   an  overdensity in the  color-magnitude
plane consistent with   a red sequence of  early-type  galaxies.   The
method  is  remarkably insensitive to   projection effects because the
requisite color-magnitude relation   is unlikely to  be  met by random
projections.   The enhanced sensitivity of the   CRS algorithm to real
clusters and its  robust insensitivity to projections  is a result  of
making the assumption  that all significant  real clusters have a  red
sequence  of early-type galaxies.  Though   this assumption must break
down at some point in the early formation  of a cluster's galaxies, it
is  observed to hold  to at least $z\sim1.3$ in  a  variety of cluster
environments and a range of cluster masses, as detailed below.

\section{Observational Constraints on the Red Sequence as a Cluster Marker} 
\paragraph{}
Observations of  the red sequence  in clusters generally indicate that
the bulk of the  stars in the  cluster early-type galaxies formed at a
high redshift. Two  major lines of  evidence in  support  of this idea
have  been pursued by  numerous  authors.   One  is based  on  various
analyses of the color evolution of the early-type galaxies.  The other
is  based on analyses  of  the luminosity evolution  of the early-type
galaxies.  Each general   class  of analysis  is  discussed separately
below.   Furthermore, since the observational data  at  $z>1$ are much
less complete, the apparent properties of  clusters at these redshifts
are  discussed separately. All  current   data indicate that the   red
sequence is a universal signature of galaxy clusters.

\subsection{Scatter, Color and Slope of the Red Sequence}
\paragraph{}
\markcite{bowera}\markcite{bowerb}Bower,  Lucey,   \&   Ellis (1992a,b)
completed the first comprehensive  photometric   analysis of the   red
sequence properties in the Coma and Virgo clusters, and concluded that
the early-type galaxies in  each were  indistinguishable.  Considering
that   the   Coma and  Virgo clusters  are   prototypical  clusters of
completely  different   types   -- Coma   is  rich,  massive, centrally
concentrated, elliptical   rich,  and  X-ray  luminous;  and  Virgo  is
relatively poor,  much less massive,  irregular and spiral rich -- this
is a striking result. This  basic result has been  extended to a  much
larger   sample   in  the     exhaustive  low-redshift    cluster survey    of
\markcite{omaryee}LCY00.  This   multi-color imaging survey of  45  
X-ray-selected Abell clusters provides a sample which encompasses a
wide range of optical richnesses (Abell Richness Class 0 to $>2$),
Bautz-Morgan classes (I-III), masses and Rood-Sastry classes (see
L\'{o}pez-Cruz 1997 for details of the sample). Despite this
optical heterogeneity, every cluster in the sample has a red sequence;
and the k-corrected slopes, scatters and colors of these red sequences
are indistinguishable.  At least at $z<0.2$, the red sequence is a
universal and homogeneous feature of galaxy clusters (LCY00).

\paragraph{}
Other surveys extend the conclusions of \markcite{omaryee}LCY00 to
higher redshifts.  \markcite{smail}Smail et al. (1998) studied 10
optically-selected luminous X-ray clusters at $0.22<z<0.28$, and
\markcite{felipe}Barrientos (1999) studied 8 optically-selected 
clusters at $0.39<z<0.48$.  In each sample, the clusters have red
sequences with colors and scatters which are remarkably homogeneous,
and imply a high formation redshift.  Notably,
\markcite{felipe}Barrientos (1999) suggests based on the
$z\sim0.4$ data alone, that although cluster red sequence formation as
low as $z\sim1$ is possible with appropriate arrangement of the
cosmology, red sequence metallicity and star formation synchronicity,
the most reasonable formation redshift from these data is
$z_{f}\geq$2.0.  A similar result comes from \markcite{ellis97}Ellis
et al. (1997) based on three clusters at $z\sim0.54$.

\paragraph{}
The  studies   of \markcite{felipe}Barrientos (1999),  \markcite{omaryee}LCY00   and
\markcite{smail}Smail et al. (1998) are  the most comprehensive photometric studies
of  heterogeneous   clusters   samples   over    restricted   redshift
ranges. Most other   surveys analyze clusters  over  a larger redshift
range, up to redshifts of $z\sim1$.   Analyses of the evolution of
the  color    (\markcite{alfonso}Arag\'{o}n-Salamanca et  al.    1993;
\markcite{rakos}Rakos \&  Schombert  1995; \markcite{stanford98}SED98;
\markcite{kodamacolor}Kodama et al. 1998) all point to a high redshift
of formation for the red sequence stellar populations.  Complementary
analyses of the scatter (\markcite{stanford98}SED98) and evolution of
the slope (\markcite{gladders}Gladders et al.  1998;
\markcite{stanford98}SED98) over large redshift baselines indicate
similar conclusions.  The dominant uncertainty in much of this
analysis is the implicit identification of the higher redshift
clusters with lower redshift clusters of similar richness.  As pointed
out by Kauffmann (1995), in a hierarchical universe this
identification is likely not correct, as the high redshift systems
correspond to yet more massive clusters in the present day, and have a
more vigorous merger history immediately prior to the epoch of
observation when compared to present day clusters of similar mass.
However, for our purposes, it is sufficient to recognize that the red
sequences of the highest redshift clusters in these $z<1$ samples all
satisfy the photometric properties of a high formation redshift.

\paragraph{}
A final line  of evidence for the universality  of the red sequence in
clusters comes from the spectroscopic and photometric analysis of nine
$0.6<z<0.9$   optically      selected     cluster      candidates   by
\markcite{oke1}Oke, Postman, \&  Lubin (1998).  Of the nine candidates,
spectroscopic data indicate  that six are real  clusters.   Of the six
apparently real clusters, detailed information is available for three.
The first,  CL0023+0423,  is in fact the  projection  of two
sub-            Abell               Richness         Class           0 groups
(\markcite{lubin}Lubin,  Postman,  \& Oke   1998).  CL0023+0423  shows a
galaxy morphology mix  similar to the field (\markcite{lubin1}Lubin et
al. 1998), though 6 of  the 24 spectroscopically confirmed members  of
both components of CL0023+0423  have the colors of cluster ellipticals
at that redshift (\markcite{post1}Postman, Lubin,   \& Oke 1998).   The
second,  CL1604+4304, is  a massive,
centrally concentrated cluster, and shows a strong, apparently old red
sequence (\markcite{stanford98}SED98).  The third, CL1324+3011,   also
shows a prominent red sequence (\markcite{gladders}Gladders  et al.  1998).
Note that the well-formed clusters both  show a red sequence, and even
the projected groups show some evidence  of red galaxies. In addition,
another candidate, CL0231+0048,  was initially included in  the sample
precisely because it showed an anomalous lack  of a red sequence
(J.B. Oke, private comm.).  Despite being a significant overdensity on
the sky;  spectroscopic analysis shows   that it is not a real  cluster
(\markcite{oke1}Oke et al. 1998).   Furthermore, the fact that the Oke
et al.  sample includes  cluster candidates  which have a  broad
range of galaxy mixtures again argues against suggestions that optical
cluster candidate catalogs such as  these are biased to only  clusters
with  old,  well-developed galaxy  populations.   Rather,  the optical
samples available to  date appear to sample   a wide range of  cluster
types, and   all true clusters  in the  samples  appear to   have a red
sequence.

\subsection{The Redshift Evolution of the Red Sequence Luminosity}
\paragraph{}
The measured  luminosity   evolution of cluster  early-type   galaxies
provides further evidence of the high formation redshift for the stars
in these galaxies.  Numerous fundamental-plane  studies of clusters at
redshifts up to  0.83 have demonstrated  a change in the mass-to-light
ratios of cluster galaxies consistent with that expected from passive
evolution  of   stellar  populations formed   at   high redshift (e.g.,
\markcite{vandokkum2}van      Dokkum        \&       Franx       1996;
\markcite{kelson}Kelson et al. 1997;
\markcite{bender1}Bender et  al.  1998;
\markcite{vandokkum1}van         Dokkum et   al.                1998;
\markcite{jorg}J\makebox[0pt][l]{/}orgensen et al.       1999).
\markcite{felipe1}Barrientos,   Schade,    \&   L\'{o}pez-Cruz  (1996),
\markcite{schade1}Schade et  al.   (1996) and \markcite{schade}Schade,
Barrientos, \& L\'{o}pez-Cruz (1997) demonstrated similar evolution in
the $M_B-\log{R_e}$ relation (a projection of the fundamental plane);
as did \markcite{pahre}Pahre, Djorgovski, \& de Carvalho (1996),
\markcite{barger}Barger et al.  (1998),  and
\markcite{zeigler}Ziegler et al.  (1999) using the evolution of the
Kormendy relation (\markcite{korm}Kormendy 1977).  Further evidence
for luminosity evolution comes from the evolution of the
Mg$_b$-$\sigma$ relation (e.g., \markcite{bender}Bender, Ziegler, \&
Bruzual 1996; \markcite{ziegler1}Ziegler \& Bender 1997).  Taken
together, these studies provide convincing evidence of the essentially
passive evolution of the stellar populations in cluster ellipticals.
\markcite{zeigler}Ziegler et  al.   (1999) do note  that there   are a
number of possible uncertainties in such studies -- however, the general
conclusion that the bulk  of the stellar  population in these galaxies
must have formed at $z_f>2$ still holds.  As in  the case of photometric
color-based studies, it is difficult to provide more stringent limits,
as this   requires observations of  clusters at  higher redshifts than
those explored so far.

\subsection{The Red Sequence at $z>1$}
\paragraph{}
The universality of the red sequence is not yet well tested at $z>1$,
and analysis in the context of galaxy formation models is quite
limited (e.g., \markcite{stanford98}SED98).  This is simply due to a
lack of candidate clusters at these redshifts, and the observational
cost required to observe them.  Despite this, optical and IR imaging
of a portion of the limited sample of clusters does point to the
universal presence of the red sequence.  The highest redshift cluster
discovered to date in a blank field imaging survey is CIG J0848+4453
at $z=1.273$ (\markcite{stanhighz}Stanford et al.  1997).  This
cluster was first detected as a spatially compact overdensity of red
galaxies in an optical-IR imaging survey -- essentially as a detection
of the red sequence. A nearby, possibly associated cluster, RX
J0848.9+4452, is described by \markcite{rosati1}Rosati et al. (1999)
and also shows a strong red sequence. The radio galaxy 3C324 at
$z=1.206$ is also embedded in a large cluster, which again shows a
strong red sequence (\markcite{dickinson}Dickinson 1995).  A further
interesting case is the cluster AXJ2019+112 at $z=1.01$. As discussed
in \S1, this cluster was first thought to be an optically dark cluster
(\markcite{hattori}Hattori et al.  1997), but was later revealed to be
a normal cluster at $z=1.01$ with a strong red sequence
(\markcite{benitez}Ben\'{\i}tez et al.  1999). da Costa et al. (1999)
present color-magnitude diagrams for two $z\sim1$ cluster candidates
detected in the ESO Imaging Survey (\markcite{olsen}Olsen et al.
1999); both show a red sequence of early-type galaxies at the color
appropriate to the estimated redshift, and are thus both likely real
clusters.  These candidates were selected on the basis of galaxy
overdensities using the matched-filter technique, which 
makes no assumptions about galaxy colors, and so the presence of a red
sequence is not expected {\it a priori}.  Notably, neither of these
clusters is particularly rich (M.  Franx, private comm.), which is
expected given the total area from which the candidates were selected
(1.1 deg$^{2}$, \markcite{olsen}Olsen et al. 1999).

\paragraph{}
It is striking that all the high-redshift clusters studied so far have
red sequences.  Except for CIG J0848+4453, the selection techniques
for these clusters do not in general require such a signature --
indeed, the cluster AXJ2019+112 was selected by a combination of X-ray
emission and the lensing of a background quasar
(\markcite{hattori}Hattori et al.  1997) without any reference to
cluster galaxies whatsoever.  Moreover, the richness or mass range
encompassed by these clusters is quite large (
\markcite{smail3c324}Smail \& Dickinson   1995; 
\markcite{hattori}Hattori et   al.   1997; 
\markcite{stanhighz}Stanford et al.  1997), 
with implied richnesses ranging from Abell Richness Classes 0 to 2,
and most of the clusters being relatively poor.  The implication is
that the evolution of the cluster galaxy population towards a red
sequence occurs at relatively small mass scales and thus prior to the
assembly of large clusters.  A recent weak-lensing analysis of
MS1054--03 (a massive cluster at $z=0.83$) demonstrates this as well.
MS1054--03 is clearly not relaxed, and appears to be composed of
several sub-clumps with velocity dispersions (presuming an isothermal
distribution) of $\sigma_{1}\sim650$ km s$^{-1}$
(\markcite{henk}Hoekstra, Franx, \& Kuijken 2000). Despite this, it
already has a strong red sequence of early-type galaxies.  Moreover,
HST imaging and Keck spectroscopy of MS1054--03 reveals the presence
of numerous galaxy-galaxy mergers, many of which are occurring between
galaxies which are already red (\markcite{franx}Franx et al.  1999).
Again, the implication is that the entrenchment of a red sequence in
the cluster (i.e., the formation of the bulk of the stellar population
in the eventual red sequence galaxies irrespective of the dynamical
state of the host galaxies of said stars) occurs before the formation
of the cluster as a massive relaxed system.  In support of this
general picture is the recent result of \markcite{ponman}Ponman,
Cannon, \& Navarro (1999).  On the basis of an analysis of the entropy
of the X-ray gas in clusters and groups, Ponman et al.  conclude that
some form of pre-heating is require to explain the observed trends in
X-ray temperature with cluster mass.  Ponman et al.  argue that this
requires that the galaxies in clusters must form the bulk of their
stellar populations prior to the assembly of the cluster, as is
generally expected from hierarchical structure formation models.

\paragraph{}
Present observations thus indicate a paucity of mass overdensities
(clusters, in the cosmological sense) which do not have a red sequence
of early-type galaxies, even at $z>1$.  While it still remains
possible that such clusters do exist, perhaps at some very early,
low-mass stage of cluster assembly, it seems likely that they account
for only a very small fraction of all clusters. Clearly, any search
for clusters using the CRS method will tend to be biased against such
clusters. Paper II discusses this in further detail, by examining the
effect of variations of the blue fraction on detection efficiency. In
general, it seems eminently reasonable to search for clusters using
the red sequence.

\section{CRS Implementation}
The implementation of the CRS method considers four main parameters
for each detected galaxy in a survey: positions $x$ and $y$, magnitude
$m$, and color $c$.  The color has an associated color uncertainty,
$\delta c$, and the positions and magnitude are considered to be
error-free. Photometric errors on the magnitude are ignored because
the red sequence is nearly horizontal in a color-magnitude diagram
(CMD).  From the four input parameters for a set of observed galaxies,
we wish to locate those positions which likely correspond to a galaxy
cluster, and estimate a redshift, $z$.  In summary, this is done as
follows:

\begin{quote}
1) Define a series of overlapping color slices in the $c$-$m$ plane,
guided by a fiducial red sequence model.

2) Select a subset of all galaxies as belonging to each slice, based on
on the probability (evaluated from $m$, $c$, $\delta c$ and the model)
that each galaxy belongs to that slice.

3) For a given subset, compute weights for all selected galaxies. The
weight is based both on the galaxy magnitude, and the probability that each
selected galaxy belongs to that slice.

4) Compute the weighted surface density of galaxy positions. All the
slices taken together define a form of volume density in $x$, $y$ and
redshift.

5) Identify the peaks in this volume and select
clusters candidates using some significance cut.

\end{quote}
The detailed implementation of this method is described in the following subsections.

\subsection{Color Slices}
\paragraph{}
First, the expected photometric behavior of the red sequence as a
function of redshift is determined from a default model of the red
sequence, which provides the color and slope of the red sequence as a
function of redshift. In the context of typical galaxy spectral
synthesis codes, defining the model essentially consists of selecting
a cluster formation redshift, $z_{f}$, and deducing a
luminosity-metallicity calibration so that the observed slope of the
red sequence at low redshift is reproduced (see \S5 for an example).
The details of the model are relatively unimportant, as any
application to a real survey should include a verification of the
model colors by comparison to observations of known clusters over the
redshift range of interest.  Even if the model is not particularily
well calibrated, errors in the model colors will tend to manifest as
biases in secondary derived parameters such as the redshift, and will
have a minimal effect on the $a$ $priori$ chance of detecting a given
cluster. Moreover, minor errors in the model slopes are likely to have
a minimal effect, since the slopes are small to begin with.

\paragraph{}On  the  basis    of   this   model, the
color-magnitude diagram of the entire sample is divided into a number
of regions bordered by the expected red sequence at various values of
$z$.  The choice of these bounding $z$ values is driven by the data.
Over the redshift range set by the filter combination, bounding red
sequences are placed at color separations, $\Delta c$, corresponding
to:
\begin{equation}
  \Delta c=(\delta c(c,m)^{2}+\delta RS^{2})^{\frac{1}{2}},
\end{equation}
\noindent   where $\delta c(c,m)$   is  the  measured color  error  of
galaxies at a given color and magnitude.  This fiducial color error is
evaluated at values of $c$ and $m$ which correspond to a fixed
absolute magnitude ($M^*$ is used here) at various redshifts.
Clearly, the color slices can be thought of as redshift slices, with a
width in redshift, $\Delta z$, which is a function of $\Delta c$ and
the details of how color is mapped onto redshift. The second term,
$\delta RS$, is the intrinsic scatter in color seen in cluster red
sequences at moderate redshifts.  A value of $\delta RS=0.075$ is used
here (Barrientos 1999; SED98).  Each ``color slice'' defined in this
manner thus has a width sized relative to both the expected
photometric scatter and the intrinsic scatter of the red sequence at
the chosen redshift. In practice, the color slices are actually
overlapped to ensure that no clusters are poorly sampled by being on
the border between two slices.  Figure 3 shows an example of the
construction of color slices using data from the CNOC2 Redshift Survey
patch CNOC0223+00 (\markcite{yeecnoc2}Yee et al. 2000b).  The
red-sequence model used in Figure 3 is described in detail in \S5.

\begin{figure}[h]
\figurenum{3}
\plotfiddle{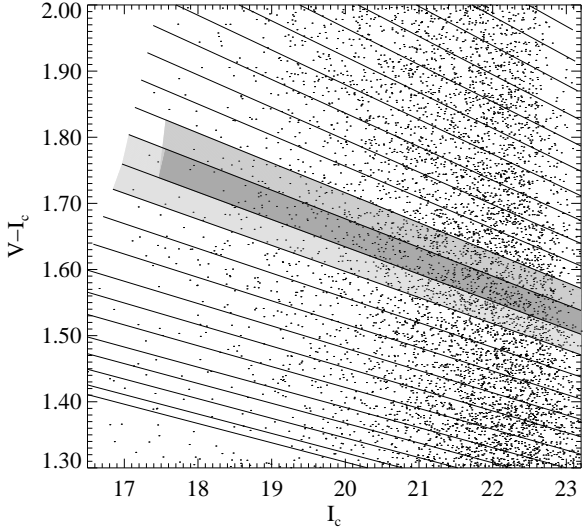}{7.2cm}{0.0}{48}{48}{-132}{0}
\caption{A portion of the observed $V-I_c$ versus $I_c$ CMD for the
CNOC2  Redshift Survey   patch CNOC0223+00.  Bounding   red sequences,
constructed as  detailed in the text, are  shown as solid lines,
from  $M^*-1$  to the survey   limits.   Two overlapping color  slices
(shaded regions)  are  highlighted, with  the  overlap region a darker
shade. For clarity, the  shading in the  second slice has been set 0.5
mags fainter than that in the first slice.  }
\end{figure}

\subsection{Galaxy Subsets}
\paragraph{}
For each color slice, defined by an upper and lower bounding red
sequence, the probability that each galaxy in the sample belongs to
that slice is then computed. This is done by assuming that the true
colors are distributed normally about the measured colors, with
Gaussian widths given by the measured color errors.  The integration
of each presumed error distribution in the desired color range then
gives the probability that any given galaxy's true color places it in
the slice. A subset of galaxies based on a probability cut is then
selected for each color slice.  A probability cut of 10\% has been
used here.  This value represents a compromise between sampling deeply
in to the available data (as all faint galaxies have low probabilities
of belonging to a given color slice because of large color errors) and
computational efficiency (which improves with the smaller number of
galaxies selected by a higher probability cut). An example of this
probability assessment, using the same data and models as in Figure 3,
is shown in Figure 4. Galaxies which pass the probability cut trace
out an envelope which is essentially as wide as the color slice at
bright magnitudes,but widens to fainter magnitudes as the size of the
color errors becomes comparable to the width of the slice. Apparent
outliers at brighter magnitudes are galaxies which have larger than
typical color errors, often due to object crowding effects.

\begin{figure}[h]
\figurenum{4}
\plotfiddle{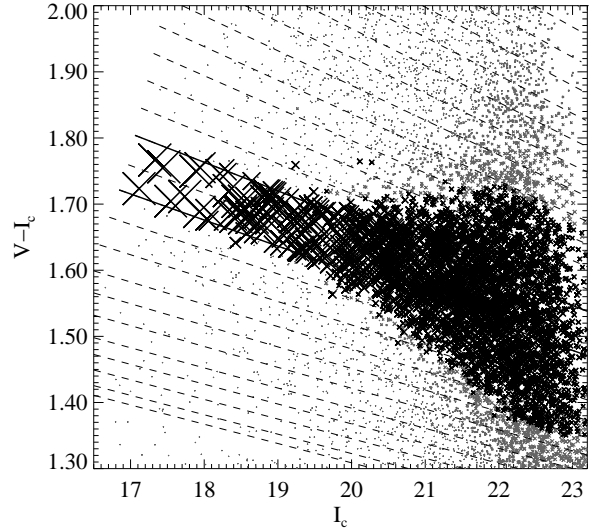}{7.5cm}{0.0}{48}{48}{-130}{0}
\caption{ 
A CMD using the same data and models as Figure 3, but
now showing the probabilities that each galaxy belongs to a particular
color slice (the bluer of the two overlapping slices shown in Figure
3). The symbol size is proportional to the probability. Darker symbols
indicate galaxies which pass the 10\% cut discussed in the text.
}
\end{figure}

\subsection{Galaxy Weights}
\paragraph{}
It is well known that field galaxy counts are in general steeper than
the faint end of the typical cluster luminosity function (e.g.,
\markcite{huan}Lin et al.  1999; L\'{o}pez-Cruz \& Yee 2000b).  Thus,
at fainter magnitudes in a given color slice the field counts will
eventually overwhelm the counts from any clusters.  Without color
selection, a similar effect happens toward brighter magnitudes, as the
cluster counts are overwhelmed by counts from lower redshift galaxies.
However, in any given color slice, because lower redshift galaxies are
excluded by the color cut, any cluster elliptical galaxies likely
dominate the bright end of a particular color slice, and so this
effect is not expected at the bright end. This recognition that
cluster galaxies of different magnitudes have a differing contrast
against the background mandates the use of magnitude-based weighting
of the galaxies, in addition to the weights taken
directly from the color-based probability assessment discussed in
\S4.2.

\paragraph{}
The relative weight which should be assigned to a galaxy of a given
magnitude can be deduced from the data, without reference to models of
the cluster and field luminosity functions. This is done by computing
the surface density of all objects in the color slice (with
probability weights based only on the colors [\S4.2], and using the
density estimator described in \S4.4 below).  Galaxies located at the
most significant peaks in the density are then designated as cluster
galaxies (with field contamination) and the rest are designated as
field galaxies.  A comparison of these two samples of galaxies, after
correcting for the field contamination in the cluster sample, then
allows one to evaluate the probability that a galaxy of a given
magnitude in the uncorrected cluster sample is in fact a cluster
member.  Put simply, if $N_{c}(M)$ is the field-corrected,
area-normalized counts for the cluster galaxies and $N_{f}(M)$ is
similarly for the field galaxies, then the probability that a galaxy
of a given absolute magnitude $M$ in the cluster sample is in fact a
cluster member is:
\begin{equation}
  P(M)=\frac{N_{c}(M)}{N_{c}(M)+N_{f}(M)}.
\end{equation}
Note that the early-type galaxy model is used to compute the evolved
value of M$^*$, with the measured M$^*$ of LCY00 taken as the local
value. 

\paragraph{}
The construction of $P(M)$ will depend on the available
data. If each color slice contains enough data that a reliable $P(M)$
can be computed, it is best to find $P(M)$ on a slice-by-slice
basis. This then allows for variations in the relationship between
clusters and the field with color (i.e., redshift). However, in the
data discussed in \S5 (and illustrated in Figure 5) the survey area is
not large enough, and so $P(M)$ has been computed cumulatively over
the entire color range considered. Furthermore, it is clear that for a
particular cluster, $P(M)$ is not absolute, as it will scale with the
richness of that cluster. However, generally the variation in the
ratio between values of $P$ at different $M$ is small, over a factor
of a few in cluster richness counts. Furthermore, since the originally
identified cluster galaxy sample is to first order representative of
true clusters in the sample, this probability is adequate.  An
illustration of the computation of $P(M)$, using initial peaks with a
range of significances (and hence richnesses), is shown in Figure
5. It shows the run of $P(M)$ with magnitude for the data discussed in
\S5, using two different significance cuts to identify initial peaks,
and hence the ``cluster'' subsample.  The size of the two different
``cluster'' subsamples differ by an order of magnitude, in that they
represent 2\% and 20\% of all galaxies in the sample, and yet the
difference between the relative values of $P(M)$, expressed by the
slope of linear fits to the $P(M)$ vs. $M$ relation, is only a factor
of 1.5. Based on this, we use a cut of 10\% in the computations
presented in \S5, as this adequately spans the extreme values
illustrated in Figure 5. Moreover, note that in any sample $P(M)$ will
be dominated by the contribution of galaxies in the poorest systems
selected by the chosen cut, as these systems contain a much larger
fraction of all galaxies. Therefore, regardless of the survey size, or
the chosen cut, $P(M)$ is naturally weighted towards the poorer
systems, for which proper weighting is more important.

\begin{figure}[h]
\figurenum{5}
\plotfiddle{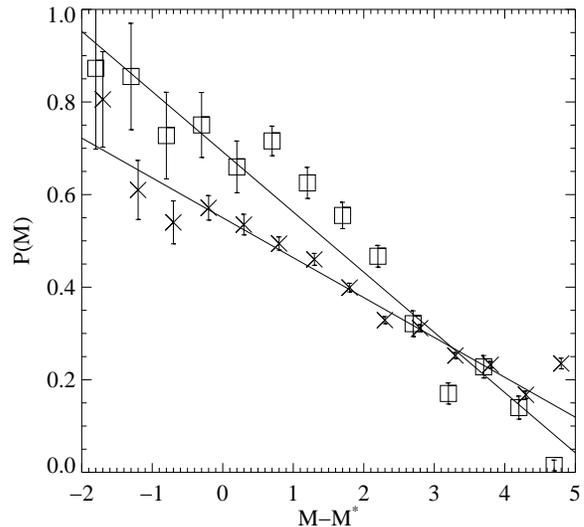}{7.5cm}{0.0}{48}{48}{-130}{0}
\caption{ 
Values of $P(M)$ versus $M$ computed for the data discussed in
\S5. Cuts representing 2\% (squares) and 20\% (crosses) of all 
galaxies in the sample are shown.  Solid lines give the best fit line
for each subsample. }
\end{figure}

\paragraph{}
An additional modification is made to $P(M)$, in which $P(M)$ for all
galaxies brighter than the expected cluster $M^*$ is set
equal to $P(M^*)$. This correction is based on the observation
that without such a modification, an unacceptable level of stochastic
noise is introduced from the few bright galaxies in a given slice.
With a final $P(M)$ computed, a weight, $w$, is then assigned to each
galaxy in a given subset.  The assigned weight is simply the product
of the probability of being included in the color-slice subset (subject to some
lower threshold as detailed in
\S4.2) and $P(M)$.

\subsection{The Galaxy Surface Density}
\paragraph{}
With a list of positions, $x$ and $y$, and associated weights, $w$,
now in hand, the next step is to evaluate the surface density of
objects. We have considered a number of estimators for this, and have
settled on simple fixed-kernel smoothing.  We considered in detail the
more complicated adaptive kernel smoothing used by \markcite{gal}Gal
et al.  (2000), but have decided that this estimator is actually
inappropriate for this particular application.  Fundamentally, the
tendency of the adaptive kernel to break up regions of high density
(i.e., clusters in the data) into smaller features may not be desirable.
Estimators such as the adaptive kernel are motivated on
mathematical grounds (\markcite{vio}Vio et al.  1994) and may not be
appropriate to every single physical situation, as is the case
here. Though appealing in general, we have found that the benefit of
the adaptive smoothing is outweighed by the additional computational
cost in this case.

\paragraph{}
The fixed-kernel estimator is used to produce a surface density map of
the galaxies in the color slice. This map is simply the weighted
convolution of the kernel with the data points at positions $x$, $y$,
evaluated over a grid spanning the range of the data (see Vio et
al. for details). Hereafter, the value of the kernel-smoothed density
at a particular location $i,j$ in the map is referred to as
$\delta_{ij}$. Two different forms were considered for the smoothing
kernel: $k(r)\propto e^{Cr}$ and $k(r)\propto e^{Cr^{1/2}}$. When
appropriately scaled, both give profiles akin to several standard
forms used to describe cluster radial density profiles.  However,
experiments on data from the CNOC2 project (see
\S5) showed the former kernel to be better suited, as the cuspy core
of the latter kernel tends to produce density maps with spurious
strong peaks associated with individual bright (and hence large
weight) galaxies which lie at the kernel cusp.  The final form of the
kernel used is:
\begin{equation}
   k(r)=Ae^{(-1.965r)},
\end{equation}
\noindent
where $r$ is the proper distance from the kernel center. In this
normalization, the radius $r$ is expressed in units of the scale
radius of the more familiar \markcite{nfw}Navarro, Frenk, \& White
(1997) (NFW) profile. The exponential constant of --1.965 is derived by
requiring that the adopted profile fit the NFW profile at intermediate
radii, and is constant provided that $r$ is expressed in units of the
NFW scale radius. The relation between the two profiles considered and
several other standard forms for cluster density profiles is given in
Figure 6. The various data for the standard profiles are taken from
Adami et al. (1998). Also, in Equation 3, $A$ is a normalizing factor chosen
so that:
\begin{equation}
   \int_{0}^{r_{max}}k(r)dr=1.
\end{equation}
A scale radius of 0.33 $h^{-1}$ Mpc has been used for the tests
described in \S5 below, a value suggested by measurements of the CNOC1
cluster sample (\markcite{carlberg1}\markcite{carlberg3}Carlberg et al. 1996, 1997b). The
maximum radius, $r_{max}$, is taken as four times the scale radius. It
should be noted that the inability of other authors to strongly
distinguish between possible profile shapes (e.g., King, Hubble or
NFW \markcite{adami1}[Adami et al. 1998]; Hernquist or NFW
\markcite{carlberg4}[Carlberg, Yee, \& Ellingson 1997])  even with large 
cluster samples indicates that the exact shape of the kernel is not of
overwhelming significance.
\markcite{lubin2}Lubin \& Postman (1996) similarly found that the
precise shape of the radial filter in the PDCS had little effect on
the detectability of clusters in that survey.

\begin{figure}[h]
\figurenum{6}
\plotfiddle{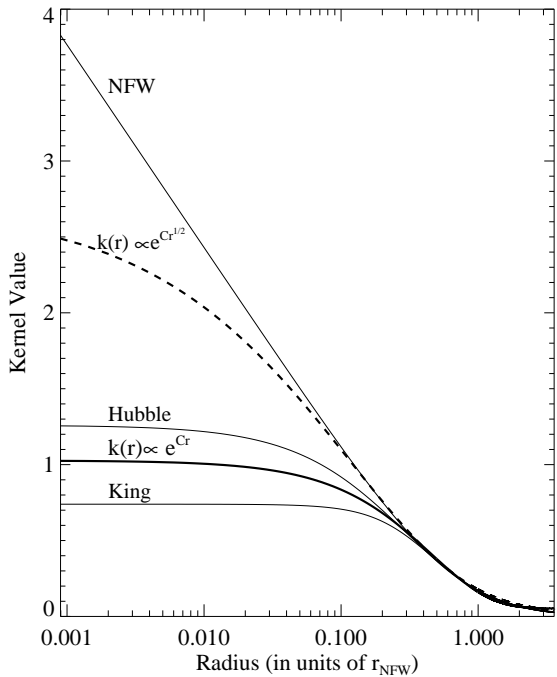}{9.0cm}{0.0}{48}{48}{-135}{0}
\caption{Various possible density-estimation
kernels. Three standard cluster profiles (NFW, King, and Hubble) as
well as the two kernels discussed in the text, are shown at radii from
0.001 to 4.0 times the NFW core radius.}
\end{figure}

\paragraph{}
Prior to the merging of the individual 2D density maps into a data
cube, the $\delta_{ij}$s must be transformed to some standard measure
(such as $\sigma$ of detection).  This is necessary because the
meaning of a given density value changes between the slices (both
because of the apparent change in the size of the fixed-metric kernel,
and because of the change in the mean density of objects relative to
the kernel size). As in Gal et al. (2000), the bootstrap technique is
used to evaluate the significance of peaks relative to the background
distribution. However, it is worth noting that a simple application of
the bootstrap technique is likely not correct, as one of the
fundamental underpinnings of the bootstrap process is the assumption
that the data points to be analyzed are independently and identically
distributed (Press et al. 1992). In data presumed to contain clusters
this is clearly not true. However, as the clusters occupy only a small
fraction of the total surface area of any map, it is possible to
exclude some fraction of the highest valued regions (and lowest, to
preserve symmetry) of the map. The same regions in the bootstrap
realization maps of the data (acquired by sampling the data with
replacement and re-running the kernel smoothing) are excluded in turn.
The probability distribution function (PDF) for density values in the
remaining areas of the bootstrap maps then closely matches the
background distribution of the initial map. Experimentation shows that
an exclusion cut of only the 10\% highest and lowest density values is
enough to solve this problem. Figure 7 illustrates the result of this
process, using the density map from a single color slice of the data
discussed in \S5.  Several derived background distributions of
$\delta_{ij}$ are shown. The first, and best fitting distribution
(solid line) is based on the bootstrap method. For comparison, another
distribution (dashed line), constructed by replacing the correlated
galaxy positions with random positions (but keeping the same weights
as the real data) is also shown.  Obviously, the restricted bootstrap
dataset provides a good representation of the background distribution
found in the input data, while the dataset based on random positions
does not.

\paragraph{}
The probability
of a given value of $\delta_{ij}$ occurring at random can then be
evaluated from the restricted bootstrap dataset. This is best done by
a direct analysis of the bootstrap PDF as it is often non-normal.
Each value of $\delta_{ij}$ in the input map is transformed into a
probability, $P(\delta_{ij})$, by evaluating the likelihood that a
value of $\delta_{ij}$ or greater occurs in the the restricted
bootstrap dataset. In other words:
\begin{equation}
P(\delta_{ij})=\frac{N(\delta_{bootstrap}\geq\delta_{ij})}{N(\delta_{bootstrap})}.
\end{equation}
The raw probabilities can then be transformed to more familiar
Gaussian sigmas if desired, though such a change is essentially
cosmetic.

\begin{figure}[h]
\figurenum{7}
\plotfiddle{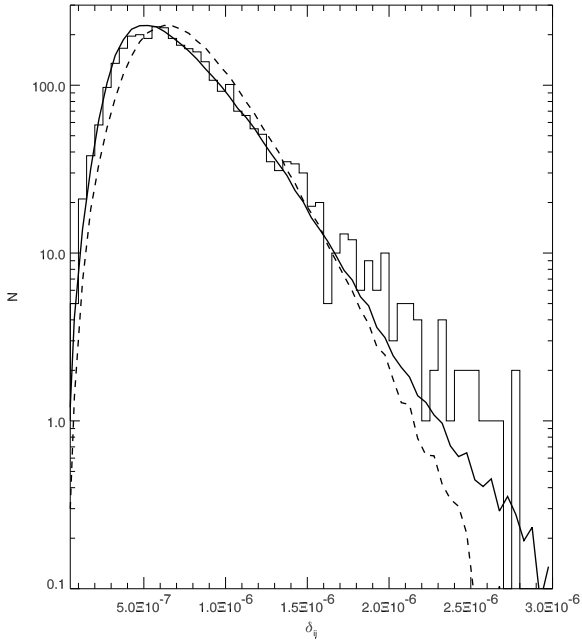}{9.0cm}{0.0}{45}{45}{-130}{0}
\caption{
The histogram illustrates the actual distribution of
$\delta_{ij}$s from a single color slice of the dataset discussed in
\S5.  The background distribution as deduced from the bootstrap
approach discussed in the text is shown as a thick solid
line. Another background distribution, based on random positions is
shown as a dashed line. The excess of high-valued points is the
signature of several significant peaks in the density map for this
particular color slice.  }
\end{figure}

\subsection{Identifying and Characterizing Clusters}
\paragraph{}
Once probability maps have been constructed for each slice, the slices
can be stacked together to form a data-cube.  The generation of a
cluster catalog from this data-cube is then simply a matter of
selecting high probability regions, subject to some predetermined
probability cut. The details of how to do this optimally will not be
described here, as the process is data dependent, though not
complicated. A further exploration of this topic, in the context of a
particular survey and detailed simulated data, can be found in paper
II (\markcite{trcs2} Gladders \& Yee 2000b), or in \S5 below.  It
should be noted that other authors have used relatively simple
algorithms in this final step with good success
(\markcite{post}Postman et al. 1996); similar simple algorithms should
work as well.
\paragraph{}
With a cluster candidate identified, the final step is to estimate a
redshift for each candidate. This is done directly from the above
data-cube, as the redshift of a given cluster candidate is encoded by
the location of the peak probability along the color axis
of the stacked density maps at the $x$-$y$ position of the candidate.

\section{A Test of the CRS Method}
\paragraph{}
The CNOC2 dataset (Yee et al. 2000b) provides an useful testbed for
cluster-finding algorithms, and has kindly been provided in advance
of publication by the rest of the CNOC2 group.  This intermediate
redshift field galaxy survey consists of 4 patches of data.  Each
patch is $\sim0.39$ deg$^2$ in size. The surveys consists of
$UBVR_cI_c$ imaging and spectroscopy for a large sample of galaxies
($\sim$40,000 in the photometric catalog to the 100\% completeness
limit of $R$=23, with $\sim$6000 spectroscopic redshifts for galaxies
typically as faint as $R$=21.5).  However, the layout of each patch is
complex, as the patches were designed to sample the correlation
function, rather than to minimize patch edges, and so are not optimal
for cluster detections. Additionally, the CNOC2 redshift catalog is sparsely
sampled, and is somewhat biased against detecting compact structures
(due to the unavoidable exclusion of an above average number of
galaxies in dense regions when using a multi-object spectrograph -- see
Yee et al. 2000b for details). Nevertheless, the availability of a large
spectroscopic sample outweighs any such considerations, and makes the
CNOC2 survey an indispensable aid in assessing the suitability of the
CRS method. Note that in considering the output of the photometric
cluster-finding process, it is necessary to make some analysis of the
CNOC2 redshift sample in order to identify real bound structures in
the redshift data. This process is complex, particularly because of
the sparseness of the redshift sampling, and the complex weighting
functions incurred by the use of a multi-object spectrograph. A global
and rigorous analysis of bound structures in the CNOC2 redshift survey
data will be presented by the CNOC2 group (Carlberg et al. 2000). This paper
uses a simplified process, described below, to provide a first-pass
comparison with the CRS method.

\paragraph{}
To test the CRS method, a dataset was constructed over the entire
CNOC2 survey area using a subset of the available photometry. Only
fields in which $I_c$ magnitudes, $V-I_c$ colors, $\Delta(V-I_c)$
color errors, and $x$ and $y$ positions are available were used, and
these provide the input to the CRS algorithm. The total area covered
by this subset is 1.48 square degrees.  Over the CNOC2
spectroscopically unbiased redshift range ($0.12<z<0.55$), this area
corresponds to a volume of $2.7\times10^5$ $h^{-3}$~Mpc$^3$. Note that
rich clusters occur with a density of about 1 every $10^5-10^6$
$h^{-3}$~Mpc$^3$ (e.g., \markcite{bramel}Bramel, Nichol \& Pope 2000), so we
expect to find only poor clusters and groups in this volume.  The CRS
algorithm was then run across all four patches, with the redshift
range restricted to $0.1<z<0.6$ (somewhat larger than the redshift
range over which spectroscopic confirmation of the clusters can be
expected).  The red-sequence model used is based on a 0.5 Gyr burst
(drawn from the GISSEL library, \markcite{gissel}Bruzual \& Charlot
1993) completed at z=3, evolving to lower redshifts with a
metallicity-luminosity relationship which reproduces the red-sequence
slope in the Coma cluster at $z\sim0$ (\markcite{gladders}Gladders et
al.  1998).  A portion of the results of the CRS cluster-finding is
shown in Figure 8 for the CNOC0223+00 patch.  A number of significant
sources are obviously detected. In these data values of $\delta_{ij}$
have been re-cast in Gaussian sigmas, referred to hereafter as
$\sigma_{ij}$.

\paragraph{}
Individual sources were then selected from the data-cubes using the
clump finding algorithm of Williams, de Geus, \& Blitz (1994). This
algorithm was originally developed for locating structure in
position-position-velocity radio data.  Williams et al.  discuss the
accuracy of their clump finding technique in detail; in summary, it
works by contouring the data in 3-D at multiple levels, and following
clumps through the various levels from an initial peak value, identify
sub-clumps as appropriate. Because the algorithm assumes no clump
profile, it is readily applied to the CRS position-position-redshift
data-cubes. Via extensive simulations, Williams et al.
recommend that the data be contoured at levels which are separated by
$\Delta T$=$2T_{RMS}$.  In the original application, $T_{RMS}$ is the
root-mean-square error in the temperature of the
position-position-velocity data.  A similar statistic can be estimated
from the CRS data-cubes by examining random realizations of the
position and weight data used to construct the data-cubes. These
random realizations, which are distinct from the bootstrap
realizations discussed in \S4.4, have the same weights as the original
input data but randomized positions, and are used within the algorithm
primarily to account for irregular patch edges.  In the particular
CNOC2 data discussed here, the RMS of the random realizations is about
$\Delta\sigma_{ij}$=0.55. The RMS is not directly indicated by 
the distribution of $\sigma_{ij}$s in the real data-cube as a whole
(i.e.  the bootstrap normalized data-cube, cast in $\sigma_{ij}$s, does
$not$ have a noise of $\Delta\sigma_{ij}$=1), as this represents a
scaling relative to the distribution of the data as a whole, including
all real structures.  Hence, for the data discussed here, the Williams
et al. clump-finder was run using $\Delta \sigma$=1.1, down to a
contour level of $\sigma_{ij}=2.4$. This means that the lowest
significance peaks detected have a maximum significance of 3.5$\sigma$
(1 contour above the lowest contour level) and all detected peaks,
regardless of maximum significance, are traced to 2.4$\sigma$.

\paragraph{} For each selected clump in the data-cube, a redshift was
estimated from the position of the clump maximum projected along the
$z$-axis of the data-cube. All CNOC2 galaxies with spectroscopic
redshifts within a projected distance of 0.5 $h^{-1}$ Mpc of the
projected clump in $x$ and $y$ were then selected.  A simple
friends-of-friends analysis of these galaxies (compared to the entire
redshift sample) was then performed.  A fixed linking length of 0.5
$h^{-1}$ Mpc in angular space, and 400 km~s$^{-1}$ in redshift space
was used throughout.  This simple analysis serves to identify likely
associated objects in the redshift sample. A visual examination of the
redshift distribution of the selected subset of galaxies, taking into
account both the distance from the clump center and the number of
friends each galaxy has is then used to select (if possible) a
structure corresponding to the identified clump. In cases where more
than one structure appears in the redshift subsample, the structure
closest to the $x$-$y$ position is used.  In such cases, there are
often other identified clumps which appear associated with the other
structures, but untangling the correspondence between various clumps
and redshift structures is straightforward.  Figure 9 shows three
examples of this process using data from a portion of the central
block of the patch CNOC0223+00 (the right hand portion of the
$\sigma_{ij}$ maps shown in Figure 8). There are three significant
peaks in this region, all in the redshift interval $0.35<z<0.45$. Each
is readily associated with a different redshift structure, despite the
close proximity of the peaks both in redshift and angle. The first two
of these example redshift structures are particularly close --
separated by only 1700 km\thinspace s$^{-1}$ in redshift, and 0.9
$h^{-1}$ Mpc projected on the sky -- yet each is clearly identified as
a separate redshift structure, and as a separate CRS peak.

\begin{figure}[h]
\figurenum{10}
\plotfiddle{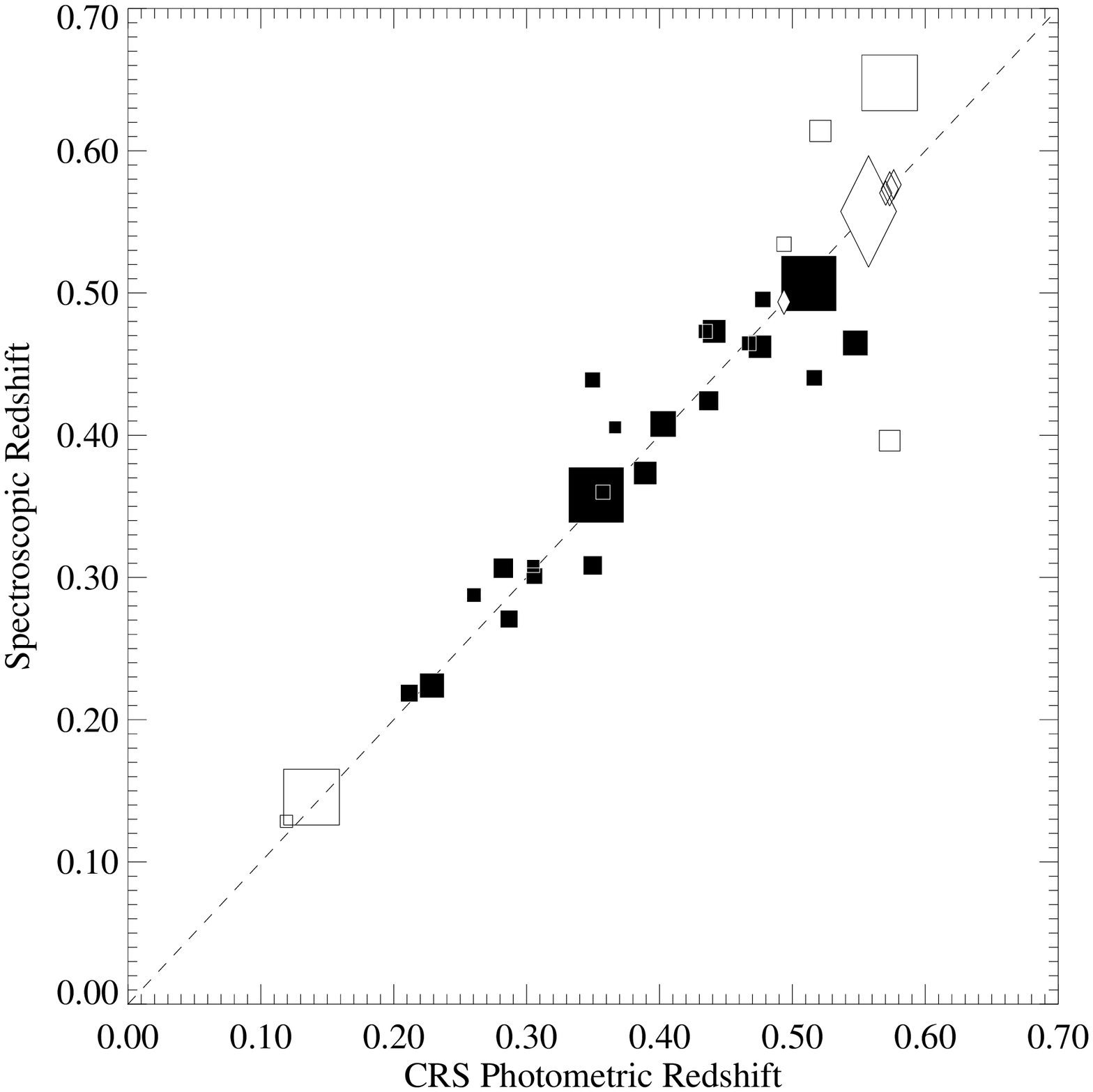}{8.0cm}{0.0}{40}{40}{-120}{0}
\caption{The comparison between the CRS photometric
redshifts (from the $I_{c}-V$ red sequence) and CNOC2 spectroscopic
redshifts.  Squares are CRS cluster candidates which have a redshift
counterpart; diamonds are those which do not.  Open symbols represent
objects in which the peak significance is in either the first or last
color slice (i.e., at the edges of the nominal redshift range of the
CNOC2 survey).  }
\end{figure}

\paragraph{} 
Figure 10 presents the comparison of the CRS algorithm output to the
CNOC2 redshift sample. CRS cluster candidates both with and without a
redshift counterpart are shown. CRS cluster candidates with peak
significance in either the first or last color slice, indicating that
the predicted redshift may be in error, are flagged. This possible
error is small at the lower redshift end (where the color errors are
small and hence color slices are narrow in redshift) but could be
quite significant at the upper redshift end. The photometric redshifts
have been adjusted by a quadratic fit to an initial comparison between
the spectroscopic and raw photometric redshifts in order to account
for any uncertainties in the early-type galaxy spectral energy
distribution model. As mentioned \S4.1, any real cluster survey is
likely to include the training data necessary for this calibration
process.  Several points of note are apparent from Figure 10.  
\paragraph{}
First, considering the subsample at $z<0.5$, the false-positive rate
is very low, with only 1 of 23 candidates not detected in the
spectroscopic sample, and even that one is at a relatively high $z$ of
$0.49$. At $z>0.5$, the apparent false-positive rate is much higher,
with 4 of 11 candidates not showing any spectroscopic counterparts.
This is likely a result of incompleteness in the redshift
sample. As the CNOC2 redshift data is sparsely sampled, very poor
cluster/group systems (with only a few bright galaxies in total), or
systems in the higher redshift range (with only a few bright galaxies
in the spectroscopic sample) may not be well represented. 

\paragraph{}
Second, the
redshift accuracy of the method is extremely good, bearing out
suggestions made in previous work (\markcite{omaryee}LCY00;
\markcite{smail}Smail et al. 1998).  The standard deviation of the
difference between the color-estimated redshifts and the spectroscopic
redshifts at $z<0.5$ is only $\Delta z=0.026$.  The apparent redshift
scatter at $z>0.5$ is much higher, with one nominally confirmed
candidate with a redshift error of 0.18. This particular candidate is
found only in the highest redshift slice (as are most unidentified
candidates), and this large redshift error may thus be due to a
misidentification of a system at a real redshift beyond the
spectroscopic reach of CNOC2 which has been falsely identified with a
lower redshift system which is along a similar line of sight and is
too poor to have been found by the CRS method as applied here. 

\paragraph{}
Regardless, the results at $z<0.5$ clearly indicate that the CRS
method is very successful in identifying real overdensities, and that
the redshift estimates are extremely accurate. There are three reasons
for the ability of the CRS method to obtain such high accuracy using
only two filters: 1) the use of a known and empirically calibrated
spectral enery distribution, 2) the fact that E/S0 galaxies have the
strongest spectral breaks and hence provide a strong color signature
with redshift, and 3) the estimation of the redshift for any given
cluster by using colors for a number of individual galaxies
simultaneously.

\paragraph{}
Finally, it should be recalled that the volume probed by the survey
considered here is quite small (in cluster terms) and that it is
expected that most of the systems identified above should be quite
poor. This expectation is well borne out by a weak-lensing examination
of several CNOC2 fields (\markcite{henk1}Hoekstra, Franx \& Kuijken
1999b) and a dynamical analysis of the groups in the CNOC2
survey (Carlberg et al. 2000). These independent analyses
indicate that most of the systems have 1-D velocity dispersions of
only 200-400 km~s$^{-1}$, well below the lower end of Abell Richness
Class 0 ($\sim650$ km\thinspace s$^{-1}$, \markcite{cnoc1global}Yee et
al.  2000a), and generally much poorer than systems identified using
previous cluster finding algorithms.  Though it is almost certain that
the CRS results shown here do not represent a 100\% complete sample at
such masses, the success in finding $any$ such low-mass systems at
these intermediate redshifts using only 2-color photometry is a strong
validation of the general method.  In addition, the extremely low
($<5$\%) false positive rate compares very favorably to
the generally higher contamination in optical cluster catalogs
constructed by other means (e.g., \markcite{oke1}Oke et
al. 1998). Furthermore, the upper limit of redshift explored here was
selected to reflect the spectroscopic limits of the CNOC2 survey;
there is no reason why the CRS method cannot be applied at yet higher
redshifts. Limiting the required wide-field data to that from
CCDs, a reasonable upper limit is about $z\sim1.4$ -- corresponding to
having the 4000\AA\thinspace break at the red end of typical CCD
sensitivities.

\section{Possible Algorithm Refinements}
\paragraph{}
A number of refinements or modifications of the general method
described above are possible.  For example, given enough information to
construct appropriate field and cluster models, one could replace the
slicing, smoothing and stacking approach used above by a
maximum-likelihood based approach (\markcite{post}Postman et al. 1996;
\markcite{kepner}Kepner et al. 1999). Such models can likely be
deduced internally from a large enough cluster-search dataset.
Additionally, entirely different density estimators could be used. For
example, the Voronoi tessellation approach of \markcite{kim}Kim et
al. (1999) is likely applicable with only minor modification of the general algorithm
described above. Many of the approaches used in X-ray astronomy to
detect diffuse collections of points embedded in a background (e.g.,
\markcite{campana}Campana et al. 1999, and references therein) are
applicable here, with the galaxies treated analogously to single
photons. The purpose of this paper is not to lay out the definitive
algorithm for using the red sequence to locate clusters. Rather, the
intention has been to describe a particular, and well-motivated
algorithm which can be used to demonstrate the effectiveness of the
basic premise of the CRS method applied to real data.
\paragraph{}
One major modification of the basic method outlined here would be the
addition of morphological information. It is possible, even with
ground based imaging, to deduce the likely morphological
classification of a galaxy using simple and computationally tractable
estimators (e.g., Abraham et al. 1994). Specifically, one can assign a
probability of being an early-type galaxy to all galaxies in a survey.
Including such analysis into the algorithm described above would be
relatively simple, as this probability can be incorporated into the
weights assigned to each galaxy. However, constructing the
morphological probability assessment of a given galaxy in real imaging
data is a significant task, as this assessment is a function of
numerous factors (i.e., the seeing, the sampling, the depth of the
imaging relative to the galaxy magnitude, the galaxy redshift and
intrinsic size etc.). Furthermore, in the CNOC2 data discussed above,
the imaging is also affected by serious camera distortions. We have
thus not attempted to incorporate morphological data into the above
analysis. However, given the basic premise of the CRS method, it is
obvious that morphological filtering will be of benefit.

\section{Conclusions}
Numerous large studies to $z\sim1$ have now demonstrated that the red
sequence of early-type galaxies is a universal and homogeneous feature
of galaxy clusters. Based on this strong observational fact, we have
developed a new technique, the CRS method, for locating clusters in
2-filter optical or infrared imaging data. Using appropriate filters,
the CRS method circumvents problems with foreground, and to some
extent background, projection effects which have been so problematic
in previous optical cluster surveys. A specific algorithm has been
constructed to implement this general method. This algorithm
constructs 2-D smoothed density maps of color slices corresponding to
redshifts bin, which are then normalized and reassembled into a 3-D
density data-cube. Peaks in this data-cube are then defined as cluster
candidates. Using data from the CNOC2 survey we have tested the CRS
method over a significant redshift range. From this test, and the
general data on the evolution of cluster early-type galaxies available
in the literature, we conclude the following:
\begin{quote}

1) The apparent universality of the red sequence in available data,
even to redshifts greater than one, indicates that the CRS method
should be applicable to the entire redshift range accessible to CCDs
($0<z<1.4$). 

2) The CRS method is a powerful new approach to finding galaxy
clusters in imaging data at optical and infrared wavelengths. The
performance of the method shows that it is able to reliably
detect rather poor bound structures ($\sigma_v \leq 400$
km~s$^{-1}$) with a very low contamination rate of less than
5\%, using data of only moderate depth.

3) The redshift accuracy of the method is $\Delta z\sim 0.025$ over
the redshift range $0.1<z<0.5$, or equivalently 10\% in redshift. This
accuracy is comparable to that achieved for single galaxies using four
(or more) filter photometric redshift methods.

\end{quote}

We have begun a large survey, the RCS, designed
to exploit the power of the CRS method (e.g., \markcite{mex1}Gladders
\& Yee 2000a). The primary survey data are 100  square degrees of imaging
in two filters ($R_c$ and $z'$), to the depth necessary to find
clusters to $z\sim1.4$. Paper II of this series (Gladders \& Yee
2000b) discusses the application of the CRS method to the RCS data
specifically, and further explores the applicability and completeness
of the method to redshifts as high as $z\sim1.4$. Unlike this paper,
the main results of paper II are based on a broad and comprehensive
set of tests performed on simulated data, created using detailed and
realistic galaxy and cluster models.

\begin{acknowledgements}
M.D.G. thanks the Natural Sciences and Engineering Research Council
(NSERC) of Canada for support via PGSA and PGSB graduate
scholarships. H.K.C.Y. thanks NSERC for support via an operating
grant. M.D.G. thanks Ray Carlberg for numerous useful discussion on
statistics and the bootstrap method. We are particularly indebted to
the CNOC2 group for providing access to the CNOC2 dataset in advance
of publication. We also thank the referee, Bob Nichol, for his
detailed comments.

\end{acknowledgements}

\newpage
\onecolumn

\begin{figure}[htb]
\figurenum{8}
\epsscale{0.9}
\plotone{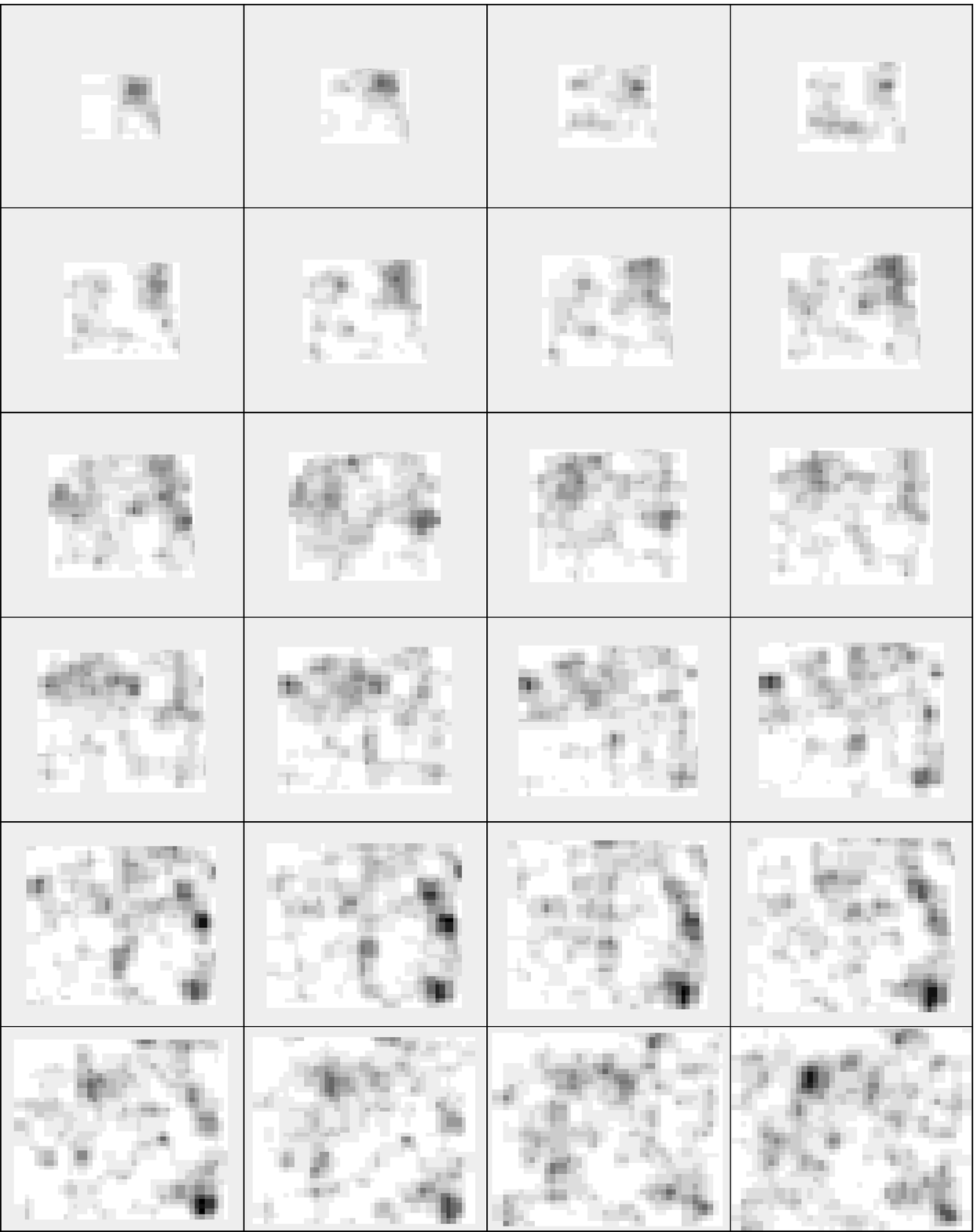}
\caption{The individual maps of $\sigma_{ij}$ from
the patch CNOC0223+00. Grayscale levels from white to black correspond
to values of $0<\sigma_{ij}<5$ in the normalized maps. The central
redshift of the slices runs from $z\sim0.1$ to $z\sim0.6$, from the
top left to bottom right. The maps are displayed at all redshifts with
the same pixel scale of 0.125 $h^{-1}$ proper Mpc. The significant
peaks in the last two rows are discussed in more detail in the text
and Figure 9.}
\end{figure}

\begin{figure}[htb]
\figurenum{9}
\epsscale{0.8}
\plotone{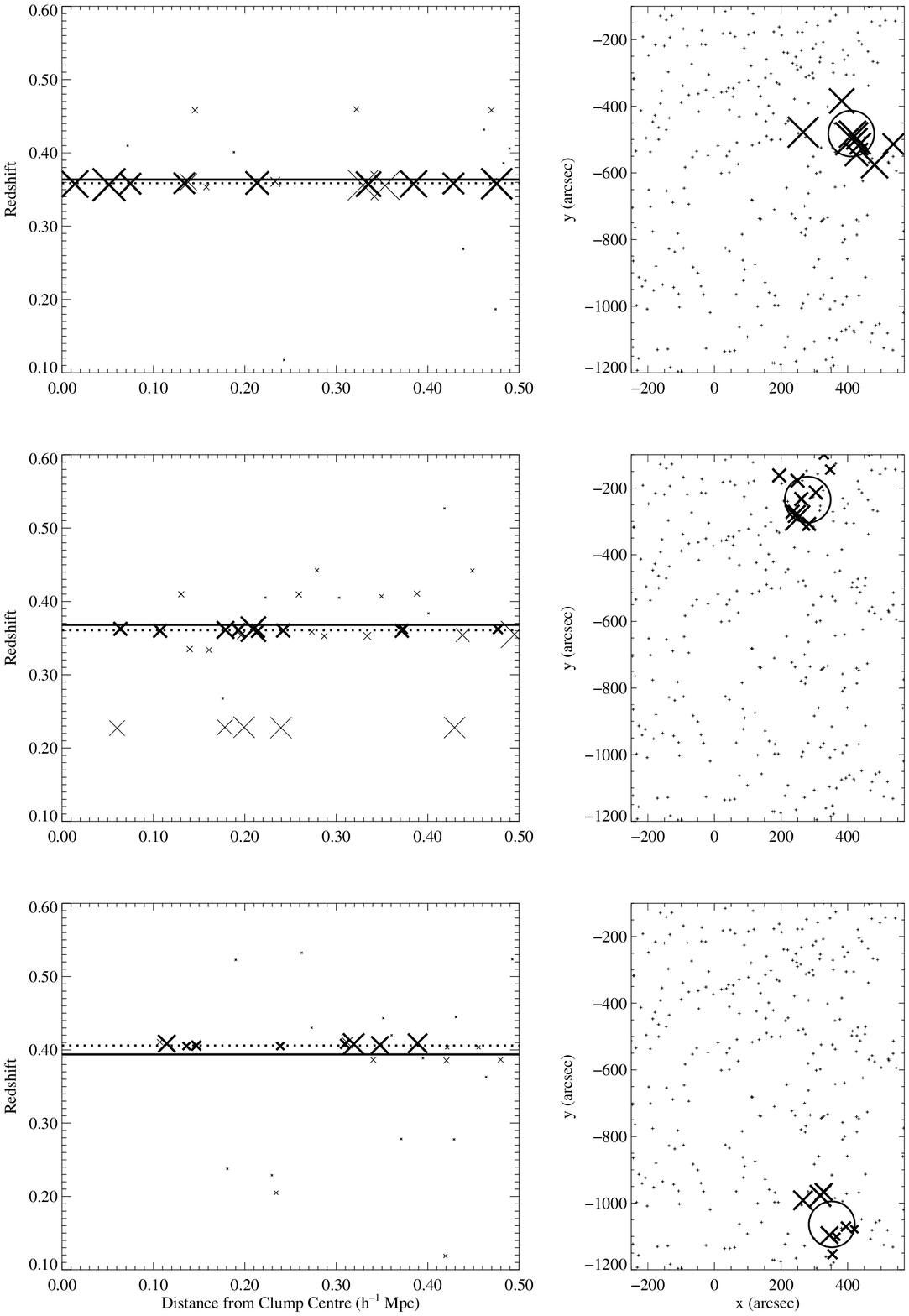}
\caption{
Example candidate clusters from the CNOC2 database found by the CRS
method and identified with real redshift structures. The right hand
panels show the $x$-$y$ plots of all galaxies in the region with
measured redshifts (small pluses, same region in each panel). Galaxies
taken to be associated with the peak under consideration are given by
large heavy crosses, with the projected $x$-$y$ position of the CRS
peak marked by a circle. The panels on the left show the redshift
vs. distance from this projected peak center for galaxies (crosses)
near each candidate. Symbol sizes are proportional to the number of
`friends' each galaxy has in the redshift sample. In each, the
predicted redshift from the CRS peak is shown as a horizontal line,
and the computed mean spectroscopic redshift of the group identified
from the redshift catalog is shown as a dashed line.  The heavy
symbols indicate galaxies taken to be associated with that
overdensity. Note that the CRS method successfully seperated two
galaxy groups with close proximity in both projected angular (4.5
arcmin, or 0.9 $h^{-1}$ Mpc) and velocity ($\sim1750$ km\thinspace
s$^{-1}$) space. }
\end{figure}

\end{document}